\documentclass{nime-alternate}
\usepackage[super]{nth}
\usepackage{graphicx}

\begin{document}

%
\conferenceinfo{NIME'19,}{June 3-6, 2019, Federal University of Rio Grande do Sul, ~~~~~~  Porto Alegre,  Brazil.}

\title{Vrengt: A Shared Body--Machine Instrument for Music--Dance Performance}

%
%
%
%
%

\numberofauthors{3} 
%

\author{
%
%
\alignauthor \c{C}a\u{g}r\i{} Erdem 
\\
       \affaddr{RITMO Centre for Interdisciplinary Studies in Rhythm, Time and Motion}\\
       \affaddr{Department of Musicology}\\
       \affaddr{University of Oslo}\\
       \email{cagri.erdem@imv.uio.no}
\alignauthor Katja Henriksen Schia 
\\
       \affaddr{PRAXIS}\\
       \affaddr{Norwegian Contemporary Dance Company}\\
       \email{katjaschia@gmail.com}
\alignauthor Alexander Refsum Jensenius 
\\
       \affaddr{RITMO Centre for Interdisciplinary Studies in Rhythm, Time and Motion}\\
       \affaddr{Department of Musicology}\\
       \affaddr{University of Oslo}\\
       \email{a.r.jensenius@imv.uio.no}
}




\maketitle
\begin{abstract}

This paper describes the process of developing a shared instrument for music--dance performance, with a particular focus on exploring the boundaries between standstill vs motion, and silence vs sound. The piece \textit{Vrengt} grew from the idea of enabling a true partnership between a musician and a dancer, developing an instrument that would allow for active co-performance. Using a participatory design approach, we worked with sonification as a tool for systematically exploring the dancer's bodily expressions. The exploration used a ``spatiotemporal matrix,'' with a particular focus on sonic microinteraction. In the final performance, two Myo armbands were used for capturing muscle activity of the arm and leg of the dancer, together with a wireless headset microphone capturing the sound of breathing. In the paper we reflect on multi-user instrument paradigms, discuss our approach to creating a shared instrument using sonification as a tool for the sound design, and reflect on the performers' subjective evaluation of the instrument.   

\end{abstract}







\section{Introduction}

In today's experimental performance scene, many musicians are exploring performance practices that approach dance, and many dancers are working with interactive music systems. A challenge in such exploration, however, is fundamentally different intentions ranging from particular embodied practices \cite{schacher:motion}. For a musician, the sound is the primary focus of attention, and the movements needed to produce the sound (the sound-producing and sound-modifying actions) are the result of that aim. For a dancer, on the other hand, the movements are the primary focus of attention, and any sonic output 
is secondary. It is therefore not surprising that the dancer in an interactive context does not intuitively render her movements into instrumental actions for active sound-making, but rather maintains her regular dance-actions influencing the sound generation in an abstract way. Similarly, the musician either takes the role of the composer without active involvement, or, as the performer enacting her own instrument.

\begin{figure}[tbp]
	\centering
		\includegraphics[width=\columnwidth]{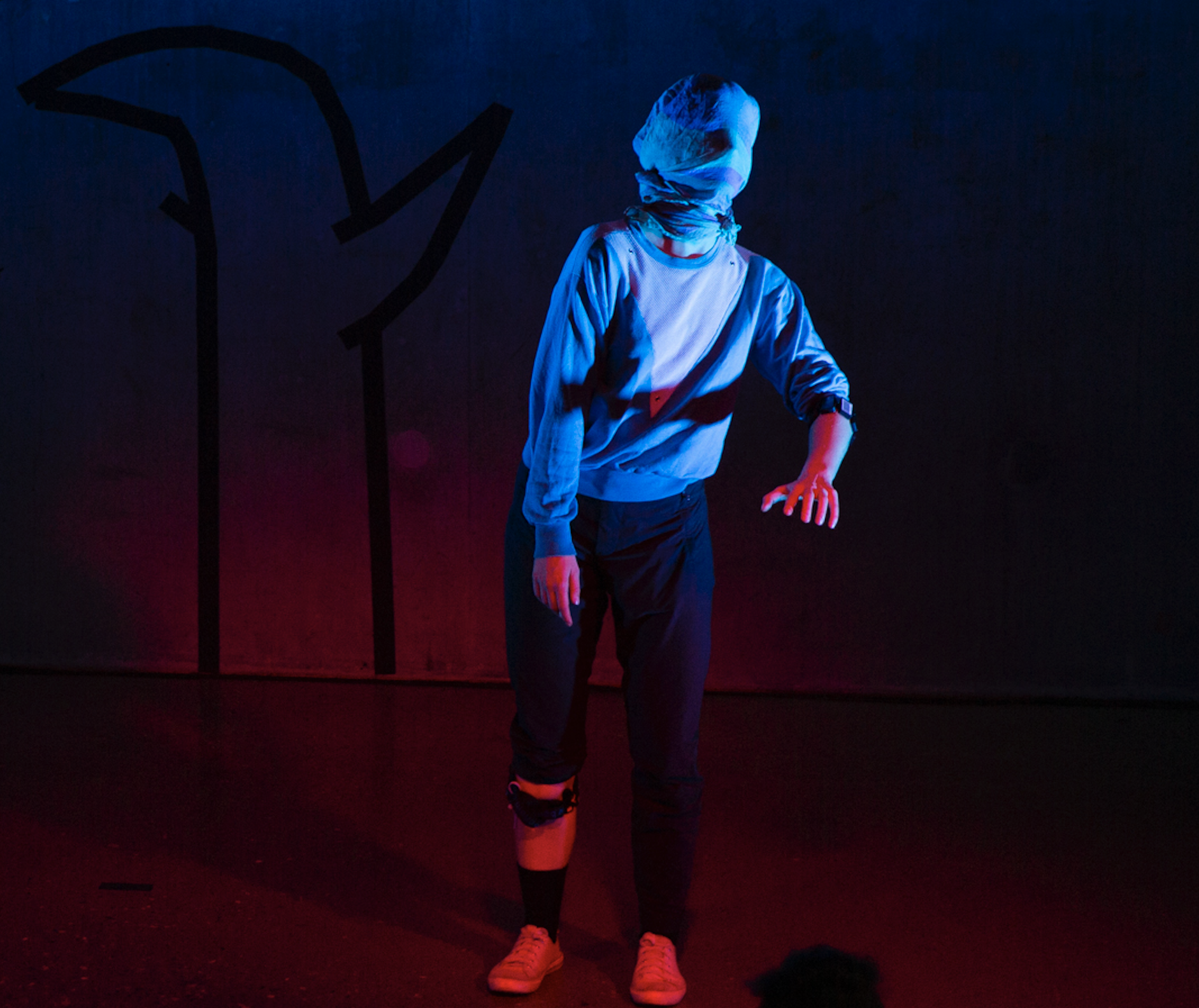}
	\caption{The dancer, blindfolded, in the first live performance of \textit{Vrengt}. (Photo: Sophie C. Barth)}
	\label{fig:dancer1}
\end{figure}

In this paper, we continue our exploration of working between dance and music, this time focusing on co-performance on a ``shared'' instrument. As opposed to creating a system for interactive dance, we wanted to develop what is experienced as one, coherent instrument that enables a true partnership for the musician and dancer. The challenge, then, is to what extent the dancer is able to adopt musical intentions on top of her movement practice, and whether the composer--performer can waive the control of performing while still ``playing together''? 

\section{Background}

\subsection{Between the conscious and the unconscious}

Experiencing the body as part of your subjective presence rather than a mere series of shapes on the stage, is described by dancers as ``being in your body'' \cite{purser2018being}. This is often the result of skill acquisition, which Dreyfus has argued is a continuum of less and less processing information at a cognitive level \cite{dreyfus2001phenomenological}. In other words, we operate more intuitively and less consciously as we gain expertise. In music, such skill acquisition is often based on proprioceptive relationships between a musician and  instrument \cite{paine:towards}. In fact, most human movement is found in the span between conscious and unconscious. That is, we unconsciously execute a number of physiological and biological processes for a single, deliberate task \cite{chi:emote}. This is something that has been explored in the context of music--dance performances under the labeling sonic microinteraction \cite{jensenius2011exploring,jensenius2017}. 

\subsection{Multi-user Instruments}\label{sec:multiuser}

Multi-user instruments have become more popular in recent years, but this is still a fairly unexplored territory. Historically, there are several examples of shared musical instrument practice, in particular in the form of four-handed piano works from the \nth{18} and \nth{19} centuries \cite{grinberg2016touch}. At that time, the shared performance allowed for forced intimacy in a social space, serving also as away of bridging the gaps of skills and social grades \cite{daub2014four}. In the \nth{20} century, experimental composers, such as John Cage and Karlheinz Stockhausen, explored the musical possibilities gained by exploiting the complex relationships between multiple users \cite{jorda2005multi}. But it was first with digital technologies that the idea of designing instruments specifically to work together on and around the same musical content took off \cite{jorda2008stage}. Some notable examples from the NIME community include the Tooka \cite{fels2002tooka} and Reactable \cite{kaltenbrunner2006reactable}, and a number of more recent web-based instruments may also be classified as multi-user. 







\subsection{Interactive Dance}


The second author is proficient in release-based training, which is a contemporary dance technique that focuses on performing tasks with least amount of muscle exertion by using the gravity \cite{lepkoff1999}. A challenge in an interactive dance context is to design an interface that allows the dancer control of the sound, but without sacrificing the existing performance technique \cite{siegel1998challenges}. It is particularly important to allow for \textit{flow} procedures, in which there is an immediate and causal feedback, yet at the same time a ``sense of discovery'' \cite{csikszentmihalyi1996flow}. For that reason we have been interested in using sonification as a tool, since it is often thought of as a more ``objective'' approach to rendering sound in response to data than more creatively based sound design \cite{hermann2011sonification}. There are numerous examples of the use of sonification in dance-related motion analysis \cite{naveda2008sonification}, dance pedagogy and education \cite{franccoise2014vocalizing, grosshauser2012wearable}, supporting the development of interactive dance pieces \cite{landry2017participatory, jensenius2011exploring} as well as assisting dancers with disabilities \cite{katan2016using, niewiadomski2018does}. 
In our case the sonification is not the end result, but rather a tool used as part of the creative process.

\section{Conceptual Design}

The main idea of \textit{Vrengt} was that of creating a body--machine instrument in which the dancer would interact with her body and the musician with a set of physical controllers. As such, it may seem as a quite normal setup for a music--dance performance, except that we did not want the dancer and musician to work in separate ``layers,'' but rather co-control the same sonic and musical parameters. This was conceptually different than they had done before. 
The development was done using a participatory design approach, combining a series of analyses, conversations, recording sessions, and subjective evaluations during the development of the instrument and final performance. As such, the entire process was very integrated, and both the musician and dancer felt a complete ownership of the final ``product.'' 


\subsection{Interaction Concept}

Our project grew from the concept of human micromotion, the tiniest producible and observable motion. These can be used in sonic microinteraction, which are found in most performances on acoustic instruments, but arguably not so often in digital musical instruments \cite{jensenius2017sonic}. We start from capturing the ``smallest components'' of the dancer's bodily exertions in the form of muscle signals and breathing, explore them through sonification, and then gradually build the entire system up from there.


Electromyogram (EMG) is a complex signal that represents the electrical currents generated during neuromuscular activities. It is able to report little or non-visible ``inputs'' (\textit{intentions}), which may not always result in overt body movements \cite{tanaka2015intention}. EMG is therefore highly relevant for exploring involuntary micromotion. The first author has been exploring what a muscle interface can add to the existing interaction paradigms of traditional instrumentalists \cite{erdem2017biostomp}. ``Playing with muscles'' can enhance the engagement with the instrument \cite{perez2010towards}, which should be considered at the top of the design hierarchy \cite{paine2015interaction}.




\subsection{Compositional Structure}\label{sec:composition}

The performance of \emph{Vrengt} may be seen as a \textit{comprovisation} \cite{dudas2010comprovisation}, in which the ``composed'' aspect of the instrument and choreography provides a large amount of freedom in collectively exploring sonic interactions throughout the performance. The piece was structured in three parts: 

\begin{enumerate}
\item Breath: The first part explores the embodied sounds of the dancer. Her face is covered (Figure \ref{fig:dancer1}), which physically forces her to leverage the kinesthetic and auditory senses. She explores the creation of acoustic feedback loops based on the proximity to the speakers, and these loops are modulated and dynamically controlled by the musician. 
\item Standstill: This section exploits using micromotion in sonic microinteraction. The dancer describes standing still as ``registering `what is happening' inside my body without the need of moving, which also introduces the gravity, meditation and body-awareness.'' Even though her micromotion is barely visible, the audience gradually starts to hear the direct audification of the dancer's varying neural commands leading to muscle contraction.
\item Musicking: Both performers join the active process of music-making. With the dancer's own words, this is where she is ``accessing the musician's skills and vice versa.'' During the first two sections, the audience becomes accustomed with the improvised movement patterns; the relationship between these movements and the variations in breath patterns; how her tiniest bodily exertions ``sound'' during standstill; and finally, how these sounds evolve throughout as she gradually switches from \textit{StandStill} to \textit{Musicking}. 
\end{enumerate}




\section{Implementation}\label{sec:implementation}

\begin{figure*}[tbp]
	\centering
		\includegraphics[width=0.9\textwidth]{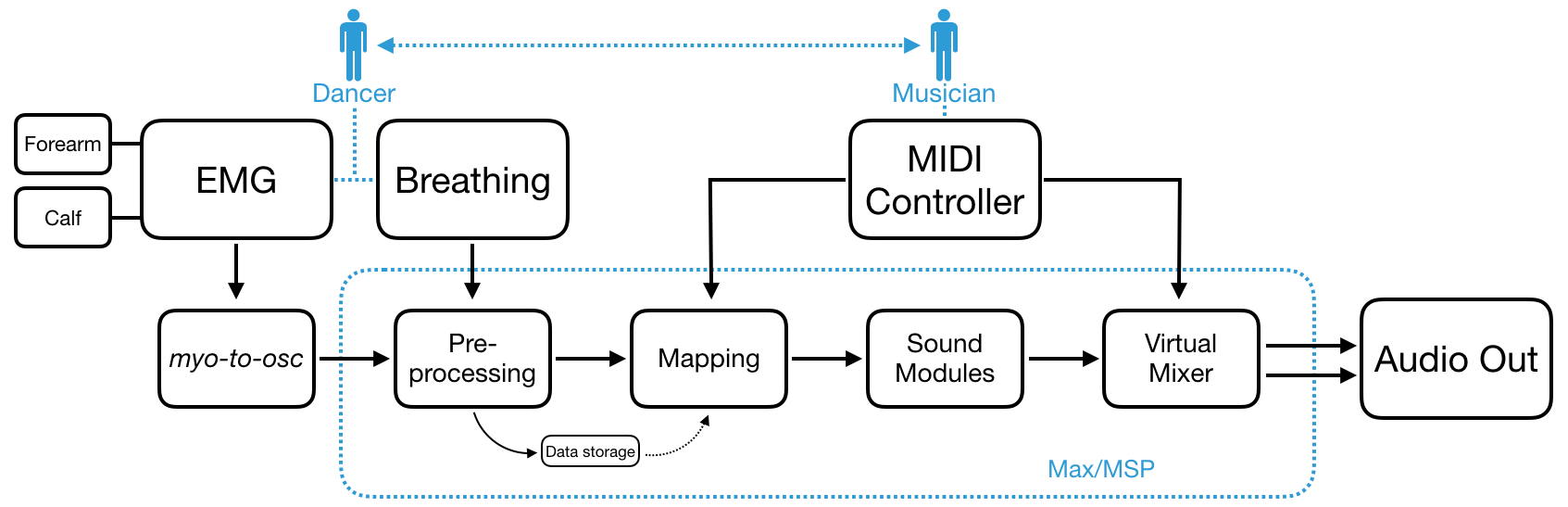}
	\caption{Signal flow diagram for the performance of \textit{Vrengt}.}
	\label{fig:diagram2}
\end{figure*}

The hardware system of \textit{Vrengt} includes  (Figure \ref{fig:diagram2}): 

\begin{itemize}
\item two Myo armbands, one placed on left forearm and one on the right calf muscle of the dancer
\item a wireless headset microphone
\item a MIDI controller
\item two laptop computers running Max/MSP patches
\end{itemize}

The armbands are connected to the computer via individual Bluetooth Low Energy (BLE) adapters for overcoming possible bandwidth limitations. The EMG data is acquired with Myo Armband's fixed sample rate at 200Hz and sent via \textit{myo-to-osc} \cite{Martin:2018ab} into Max, where the raw EMG signals are pre-processed for full-wave rectification, smoothing and feature extraction (Figure \ref{fig:diagram2}).

A lightweight and unobtrusive head-worn condenser lavalier microphone (Sennheiser SL Headmic) is used for capturing the breathing in the form of audio signals that are sent through a wireless transmitter to the laptops. 


\subsection{Mapping}

Inspired by the \emph{spatiotemporal matrix} \cite{jensenius2017sonic}, we started the mapping exploration by recording raw EMG time series of the dancer's muscle activity at different levels: \textit{micro} (during standstill), \textit{meso} (finger extensions, arm flexion) and \textit{macro} (larger actions). Then we tried various feature extractors from \cite{phinyomark2012feature}, among which we decided to use the \textit{mean absolute value} (MAV) upon a preliminary evaluation by mapping the processed data into the sound objects. After having defined the basic structure of the mappings, we subjectively evaluated each distinct action through a process of ``cross-modal interpretation.'' The dancer performed the given patterns mapped into different sound objects, and described her experiences figuratively, in order to determine the meaningful action--sound causalities (Table \ref{fig:mapping}).

Our exploration of perception--action relationships may be seen as unnecessarily time-consuming, but we found this to be necessary to better understand ``what is happening'' between the body and the sound. This is often ``arcane'' information embedded in the computational processes. The main user interface for the purpose of shared control is a custom virtual mixer that sums the individual sound modules, allowing the musician to modify the mix levels of the resultant sounds (volume, panning, effects, and so on) along with the data processes (e.g. routing and feature scaling). This is inspired by the seminal work of Alvin Lucier's \textit{Music for Solo Performer} (1965), in which his assistants controlled the sound modules throughout the performance \cite{straebel2014alvin}.

\begin{table}
\centering 
\caption{Sonic imagery of mapped relationships}
\addtolength{\tabcolsep}{-4pt}
\begin{tabular}{|c|c|c|} \hline
\texttt{Body Motion}&\texttt{Sound Object}&\texttt{Perceived Sensations}\\ \hline
Standing still & \textit{Friction}& ``Planting deeply''\\ \hline
Walking& \textit{Friction}& ``Squeaking''\\ \hline
Finger flexion& \textit{FluidFlows}& ``Squeezing a wet sponge''\\ \hline
Wrist extension& \textit{FluidFlows} & ``Casting a fishing line''\\ \hline
Abduction& \textit{Scraping} & ``Expanding like a balloon''\\ \hline
Adduction& \textit{Scraping}& ``To deflate''\\ \hline
Various& \textit{Waveshaping} & ``Shapes without images''\\ 
\hline\end{tabular}
\label{fig:mapping}
\end{table}

 \subsection{Sound Objects}

Physics-based synthesis simulates acoustic excitation and resonance features \cite{smith1991viewpoints, godoy2018sonic} to approximate responsive physical behaviors in digital domain \cite{rocchesso2003sounding}, particularly for continuous physical interaction \cite{monache2010toolkit}. In our work we have used the \textit{Sound Design Toolkit (SDT)} for physically--informed procedural sound synthesis in Max, specifically the low--level models (e.g. \textit{friction} and \textit{bubble}) and complex textures (e.g. \textit{scraping} and \textit{fluidflows}) \cite{baldan2017sound}. These have been combined with the effects processing objects (e.g. \textit{scrub{\raise.17ex\hbox{$\scriptstyle\mathtt{\sim}$}}} and \textit{pit\_shift{\raise.17ex\hbox{$\scriptstyle\mathtt{\sim}$}}}) from the \textit{PeRColate} collection \cite{trueman2001percolate}.

The \textit{SDT} basic solid interactions are based on a modular ``resonator--interactor--resonator'' structure \cite{baldan2017sound}. This allows a fairly straightforward thinking in building mapping strategies that refer to physical phenomena between objects in contact. The sound of friction, for instance, is a phenomenon that is most often present in our lives \cite{serafin2004sound}, such as the sound of a squeaking door or a knife sliding on a ceramic plate. We can then imagine several ``meaningful'' ways of associating body movements with everyday sounds. 

We used \textit{many-to-many} mappings between the calf muscle signals and the force, pressure, stiffness, dissipation and velocity parameters of the interactor algorithm (\textit{sdt.friction{\raise.17ex\hbox{$\scriptstyle\mathtt{\sim}$}}}), together with the center frequency of a narrow-Q band pass filter, to provide us with a sense of ``squeaking'' in the motions of the lower limb. Similarly, we used force, grain and velocity parameters of \textit{sdt.scraping{\raise.17ex\hbox{$\scriptstyle\mathtt{\sim}$}}} to evoke the feeling of ``filing'' when moving the upper limb. However, the perceived sense of the latter model was quite different in the end (see Table \ref{fig:mapping}). 


In liquids, sounds are heard only when the air is trapped by water \cite{leighton1994acoustic}. It is therefore a convenient approach to draw on the acoustical properties of bubbles when designing interactions with liquid sounds. A single, impulsive bubble sound is defined by its radius and rising factor (\textit{ibid}), which is simulated by exponentially decaying sinusoidal oscillators \cite{baldan2017sound}. Then, more complex phenomena can be obtained through statistical approaches as in the \textit{sdt.fluidflow{\raise.17ex\hbox{$\scriptstyle\mathtt{\sim}$}}}, which is a stochastic model. Our strategy was employing the signals of the forearm muscle to modify the speed, density and radii of a stream of bubbles, together with the amount of \textit{scrub{\raise.17ex\hbox{$\scriptstyle\mathtt{\sim}$}}} delay \cite{trueman2001percolate} for spatial enhancement. This provided us with sounds that can dynamically morph back and forth, in a continuum between rhythm and tone, echoing the \textit{unified time structuring} of Stockhausen \cite{stockhausen1989four}. 

\pagebreak

Additionally, we have explored non-linear (abstract) techniques, such as waveshape distortion, ring modulation (RM) and exponential frequency modulation (FM) for textural purposes. One technique we found intuitive, was to exponentially re-scale the sine wave carrier with multiple sine modulators in a continuous manner through several \textit{many-to-many} mappings that are also exponentially and randomly re-scaled. The result is a quasi-stochastic behavior resembling some of the non-linearities found in using extended techniques on acoustic instruments \cite{tanaka2015intention}.  

For the breath signals, we have implemented \textit{Schroeder Reverberators} \cite{PASPWEB2010} together with interconnected multiple delay lines, 
particularly for sustaining fast attacks. These are simultaneously controlled by the musician, allowing the dancer to interact with the physical space via intentional acoustic feedback loops. 

\section{Discussion}

\textit{Vrengt} has been performed twice in public so far, once on stage in a large auditorium, and another time in a club environment. In the latter it was performed together with an additional musician and a visual artist. This showed how we can use the instrument in further collaborative situations.\footnote{Video available at \url{https://youtu.be/hpECGAkaBp0}} In the following, we briefly discuss some of the thoughts we have had during those processes, specifically the subjective evaluations of the dancer and the musician.

\subsection{Musician}


For a traditionally trained musician and composer to start working with interactive dance, requires stepping outside the comfort zone. Years of experience with working within a familiar instrumental paradigm has to be exchanged with imagining oneself in the athletic and artistic circumstances of a dancer. This was the reason we decided to embark on a fairly long, exploratory journey of the dancer's movement patterns: from involuntary micromotions to deliberate full body movements. The analyses of the sensor data was followed by a number of trials during which different sound objects provided the musician with an experience-based schemata for evaluating the ecological validity of action--sound causalities (see Section \ref{sec:implementation}) and particular sound synthesis models.

The second part of the development involved rehearsals\footnote{Excerpts of video footage from rehearsals can be seen at \url{https://bit.ly/2CKl5Ia}} and verbal communication to start shaping the sound design. This phase also involved developing a shared language for describing the experience, using metaphors such as ``squeezing a wet sponge'' for grasping finger motion, or ``planting deeply'' for standing still. Such comments are necessary to understand the dancer's feelings, despite the lack of haptic experience when performing in the air. Moreover, such comments are powerful enough to define a path for future work on the relevant topics of sonic interaction design.

\begin{figure}[htbp]
	\centering
		\includegraphics[width=1\columnwidth]{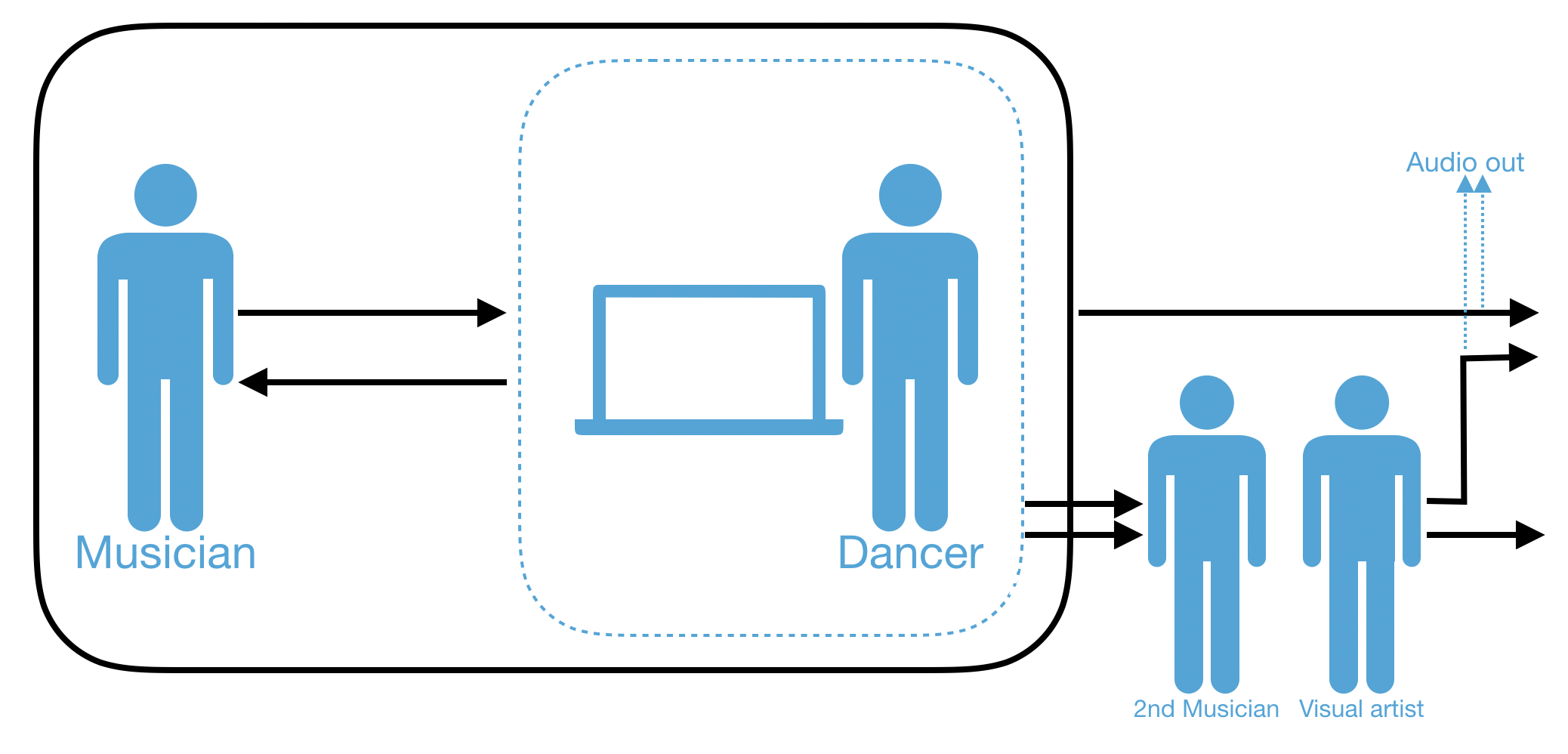}
	\caption{The setup for the final collaborative performance, showing the levels of connection between performers and instruments.}
	\label{fig:diagramM}
\end{figure}

Figure \ref{fig:diagramM} describes how the musician sees and experiences the system. The dancer is the main source of gestural input, but the musician makes the decisions of the sound objects, data scaling, and mix levels in realtime. This influences and steers the dancer who, in her own words, ``moves through listening.'' In fact, from the musician's perspective, one can draw an analogy between the dancer and the autonomous musical agents of generative systems. In this sense, the ``genericity'' of the dancer leans towards the right end of \textit{the continuum of autonomy} in \cite{tatar2018musical}, as she learns how to interact with the musician. 

The presence of the musician in this project is enacted by means of the dancer's autonomy together with the machine's data processing and sound generation abilities. This echoes the notion of ``shared control'' in the field of robotic musicianship, which often implies machine intelligence that augments human capabilities \cite{bretan2016robotic}. The purpose of such an analogy is not to get into a debate about the human versus the machine, but rather to portray the intimacy between the dancer's body and the machine, and how that is shared by the musician.




\subsection{Dancer}


From the dancer's perspective, performing with realtime sonification is fundamentally different than dancing to music. In the former case, the sonification steers the movements at both conscious and unconscious levels, and provides a  sense of coherence. However, in the latter case, you may experience a ``less or unpredictable sense of coherence.'' Throughout the collaboration, the potential of gesture--sound relationships became more clear, which allowed the dancer to develop ``a gestural repertoire and a physical landscape'' with a sophisticated control of her movement, and hence sound. This enabled listening as the main source for decision making, while intuitively moving along with ``a physical play and exploration.'' An interesting way of how she portrays her experience with the gained ability of sound-producing is as ``a duet'' of movement and the sound. As she puts it: 

\begin{quote}
    ``The precision between the muscle activation and listening drives the duet forward. It is like the ability to enter a state of not knowing where to, and how to, still with a clear sense of direction. To uncover specificity in the field of movement and sound; making sense collectively to hear the dance and to embody the sound.''
\end{quote}

One satisfactory aspect of such an instrument from the dancer's perspective, is the shift of focus from the body to the sound. This is described by the dancer as ``the sensation of moving through listening,'' which echoes Paine's \emph{techno-somatic dimension} \cite{paine2015interaction}. In addition to the ``feeling'' of playing on the instrument, she indicates how her experience with the ``sonified muscle tension'' resembles her use of tactility when an oral explanation is insufficient. 
She describes her experience of working with muscle signals as:

\begin{quote}
    ``Learning to relate to a new type of body and a new physical language that can provide an audible response.''
\end{quote}

\begin{figure}[btp]
	\centering
		\includegraphics[width=1\columnwidth]{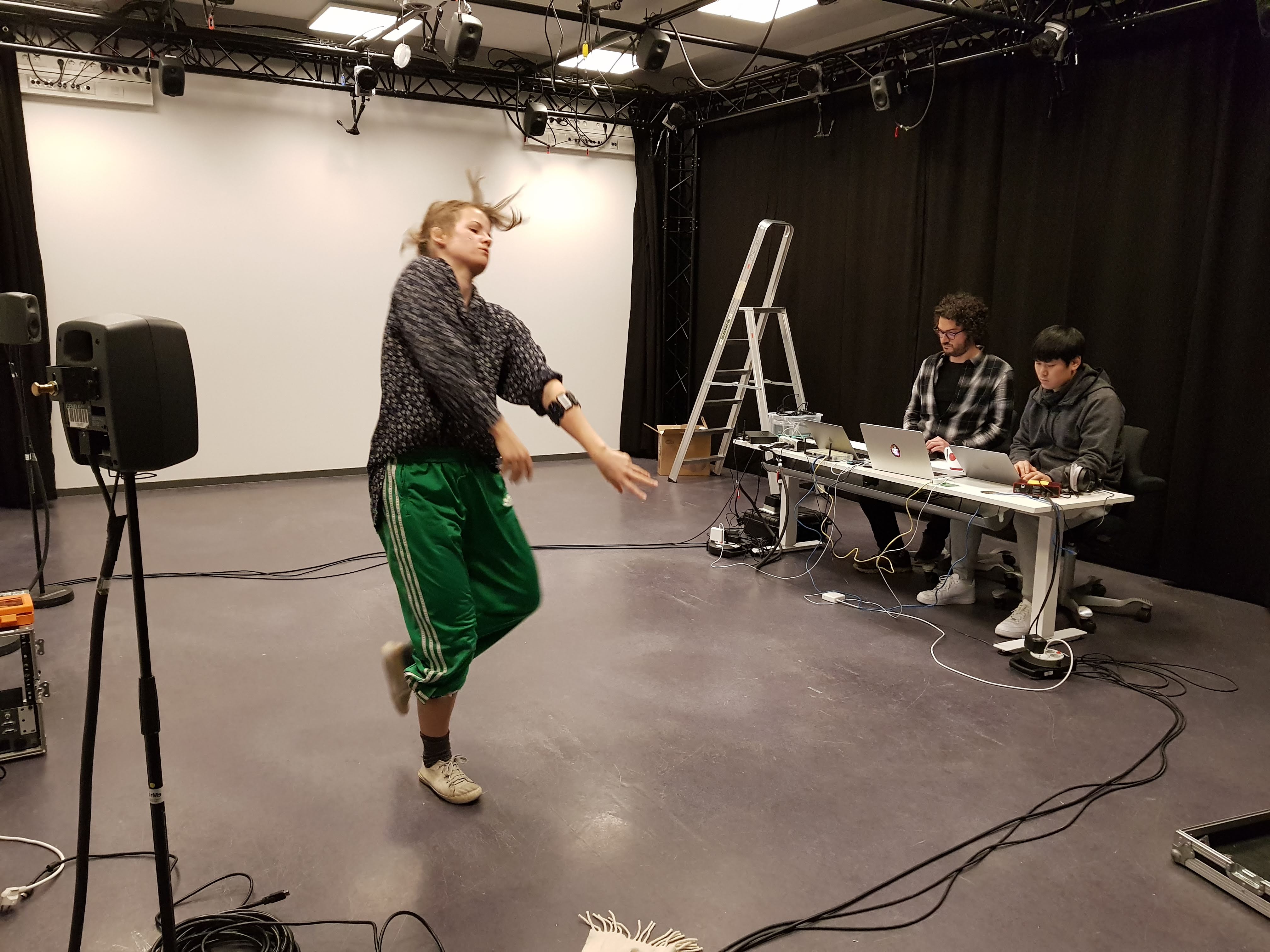}
	\caption{The dancer, the musician and the second musician in a rehearsal.}
	\label{fig:rehearsal1}
\end{figure}

\pagebreak

Her impressions about co-performing on the same instrument is described as ``playing together while accessing each other's skills.'' She uses the Norwegian word ``vrengt,'' which she exemplifies as the act of turning a sweater outwards, pointing to how the artistic intentions and skills are merged together. Furthermore, she emphasizes how each different sound object has a distinct image in her mind (see Table \ref{fig:mapping}) and she ``examined the duration, pace and consistency of every movement within them.'' Reflecting on the use of abstract algorithms for sound synthesis, she comments that they resemble shapes that she can ``fill with any image you want.'' This can be seen as opposed to more straightforward sonic imagery of physics-based models. Moreover, she cannot choose one or the other technique in terms of the level of engagement and embodied control. It is an important user-centered aspect, which should be further investigated.

\subsection{A ``Shared'' Reflection}


The usefulness of \textit{Vrengt}'s shareability to the overall aesthetics can be discussed in terms of the unity of two bodies and two machines. This relates to how Marco Donnarumma conceptualizes human-machine embodiment as ``a form of hybrid corporeality where experience, psyche, materiality and technics are always in tension against each other'' \cite{donnarumma2017beyond}. A natural outcome of this hybrid embodiment is an intimate, bodily knowledge of each other at the boundary between cognitive vs unconscious. This is different than sharing the same stage while not in a joint technological configuration. 

We observe the first aesthetic consequence of this unity in the \textit{Breath} part of the piece. What makes the role of the musician different than a tonmeister in controlling the acoustic feedback loops (see Section \ref{sec:composition}) is the multidimensional knowledge of the dancer's breath patterns. At the other end, the musician's interactions become part of how the dancer's bodily exertions happen to be in a sound-producing context. Thus, the overall aesthetics can be viewed as an \textit{n}-dimensional space of bodily and technical co-dependencies.

Similar forms of co-dependence are observed in the  \textit{Standstill} and \textit{Musicking} sections. These forms are based on the ongoing complex bodily interactions at various spatial, physiological and cognitive levels. We can then argue that the particular aesthetic results of \textit{Vrengt} would not have been achieved with other methods, such as working in separate and/or fixed layers.  

Perhaps the most significant issue in conceptualizing \textit{Vrengt} as a multi-user instrument, is the performers' uneven bodily contributions. In a more balanced scenario, the musician would use a sensor-based controller, thereby creating more of a hybrid corporeality. In our current setup, the shareability of \textit{Vrengt} is at the musician's ``fingertips'' only, when compared to the dancer's full-body experience. 

\section{Conclusions}

In this paper, we have presented the development of a multi-user instrument used in a  music--dance performance context. This project has been centered on a common apparatus, in which  shareability, sonification, micromotion, and muscle activity have been core elements. We have aimed to design a shareable instrument that blends distinct embodied skills. The final result is a joint musical expression of two performers. This has been achieved by building an entirely situated design methodology, starting from investigating the dancer's breathing and other involuntary micromotion while standing still. This was followed by using sonification as an artistic-scientific tool to explore and enhance the data in question. Furthermore, using various physics-based and abstract sound synthesis techniques allowed for subjectively evaluating their cross-modal associations and levels of embodiment. 

In future research, we will continue to build on the model of shared agency developed for \emph{Vrengt}. We are particularly interested in exploring the body as a musical interface. This will be done with a particular focus on the co-creativity of humans and machines, and using intuitive control strategies for physical modeling synthesis and embodied sonic cognition.


\section{Acknowledgments}
We would like to thank to Qichao Lan, who collaborated the project as the second musician, and Victor Evaristo Gonzalez Sanchez for his continuous support and comments throughout the development process. This work was partially supported by the Research Council of Norway (project 262762) and NordForsk (project 86892).

\bibliographystyle{abbrv}
\bibliography{vrengt}


\end{document}